\documentclass []{amsart}
\usepackage{graphicx}

\begin{document}

\title [Universal critical power for NLS] {Universal critical power for nonlinear Schr\"odinger equations with symmetric double well potential}

\author {Andrea Sacchetti}

\address {Faculty of Sciences - 
University of Modena e Reggio Emilia - 
Via Campi 213/B, I--41100 Modena, Italy}

\date {\today}

\email {Andrea.Sacchetti@unimore.it}

\begin {abstract}
Here we consider stationary states for nonlinear Schr\"odinger equations with symmetric double well potentials. \ These stationary states may bifurcate as the strength of the nonlinear term increases and we observe two different pictures depending on the value of the nonlinearity power: a simple pitch-fork bifurcation, and a couple of saddle points which unstable branches collapse in an inverse pitch-fork bifurcation. \ In this paper we show that in the semiclassical limit, or when the barrier between the two wells is large enough, the first kind of bifurcation always occurs when the nonlinearity power is less than a critical value (\ref {formula2}); in contrast, when the nonlinearity power is larger than such a critical value then we always observe the second scenario. \ The remarkable fact is that such a critical value is \emph {an universal constant} in the sense that it does not depends on the shape of the double well potential.

\end{abstract}

\maketitle

Spontaneous symmetry breaking phenomenon is a rather important effect that arises in a wide range of physical systems modeled by nonlinear equations. \ In classical physics spontaneous symmetry breaking occurs in optics, and it has been experimentally observed for laser beams in Kerr media and focusing nonlinearity \cite {Hayata,Cambournac}. \ Another natural setting in which spontaneous symmetry breaking phenomenon may arise is for Bose Einstein condensates with an effective double well formed by the combined effect of a parabolic-like trap and a periodical-like optical lattice \cite {Albiez,Raghavan,Dalfovo}. \ Also, the study of gases of pyramidal molecules, like the ammonia $NH_3$, it is a topic where spontaneous symmetry breaking phenomenon actually plays a crucial role. \ In \cite {Vardi,Jona1} has been introduced a nonlinear mean field model of a gas of pyramidal molecules; in this model spontaneous symmetry breaking explaining the presence of two asymmetrical degenerate ground states, corresponding to the different localization of the molecules, has been predicted with a full agreement with experimental data \cite {Jona1,Jona2}.

The $n$-dimensional linear Schr\"odinger equation with a symmetric potential with double well shape has stationary states of a definite even and odd-parity. \ However, the introduction of a nonlinear term (which usually models, in quantum mechanics, an interacting many-particle system) may give rise to asymmetrical states related to a spontaneous symmetry breaking effect. \ The governing equations are nonlinear Schr\"odinger equations of Gross-Pitaevskii type
\begin{eqnarray}
i \hbar {\partial_t \psi} = H_0 \psi + \epsilon |\psi |^{2\mu} \psi \, , \label {formula1}
\end{eqnarray}
where $\epsilon$ is the strength of the nonlinear term, $\mu >0$ is the nonlinearity power, and $H_0$ is the linear Hamiltonian with a symmetric double well potential. \ When $\mu=1$ we have a cubic nonlinearity and the resulting equation has been largely studied \cite {Aschbacher,Rose,Kirr,Grecchi1,Grecchi2,Sacchetti1}. \ Recently, for higher values of $\mu$ the resulting equation has been the object of an increasing interest with several interesting physical applications \cite {Gisin}. 

\begin{center}
\begin{figure}
\includegraphics[height=5cm,width=7cm]{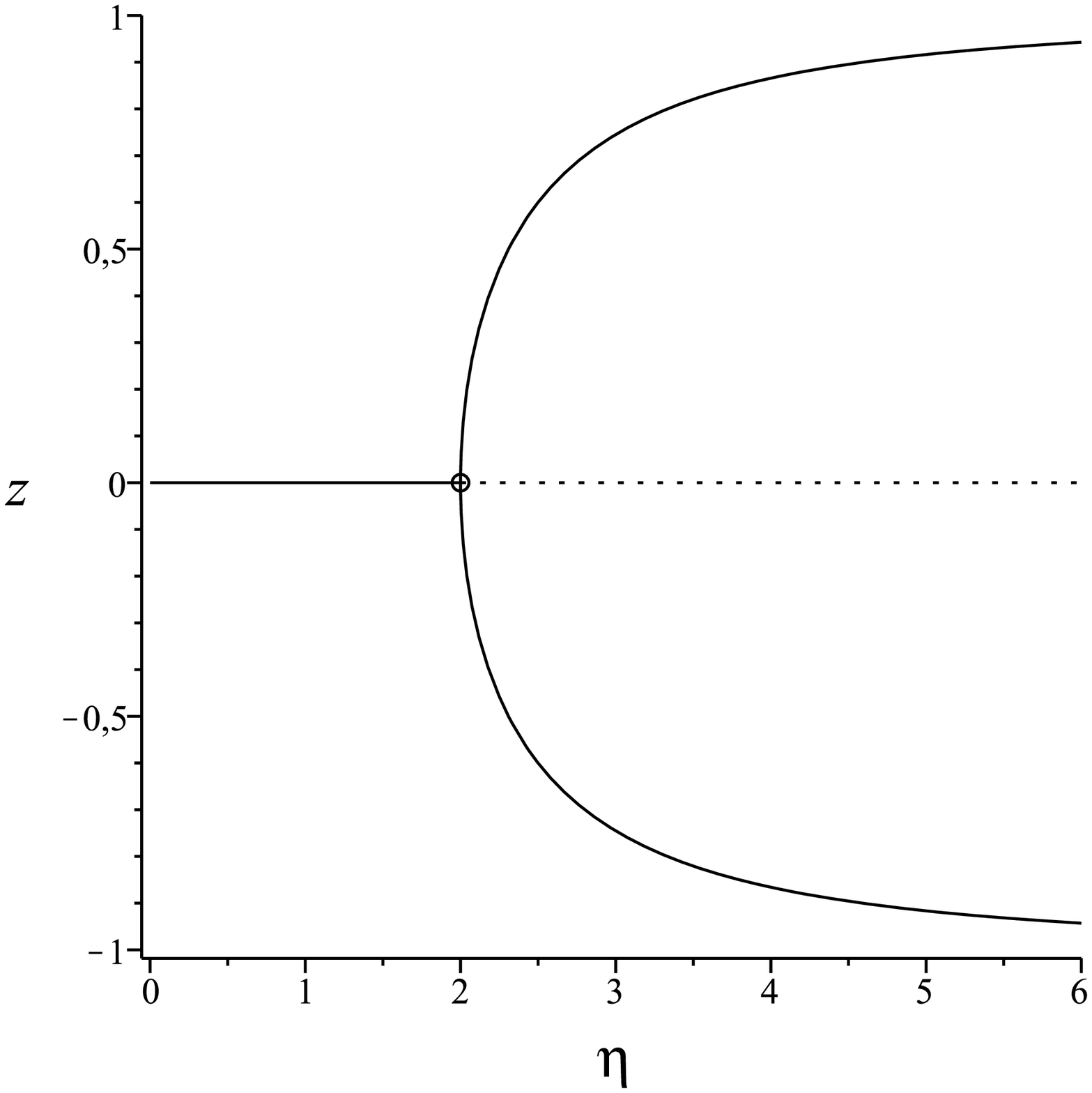}
\includegraphics[height=5cm,width=7cm]{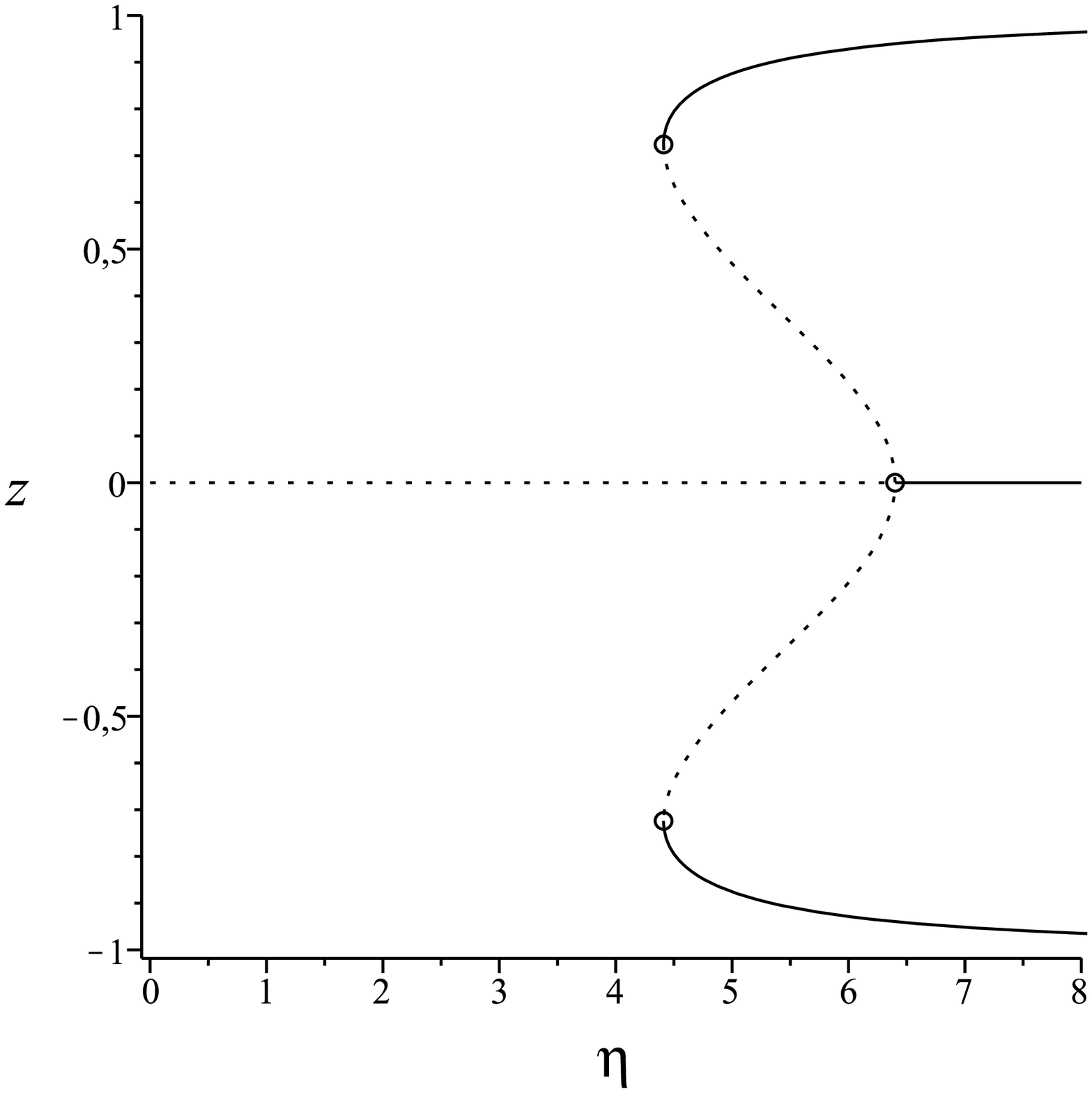}
\caption{\label {Fig1} In this figure we plot the graph of the stationary states around the minimum of the energy (full lines represent stable stationary states, broken lines represent unstable stationary states) as function of the nonlinearity parameter $\eta$ for cubic nonlinearity (i.e. for $\mu =1$) in the top panel, and for higher nonlinearity power $\mu =5$ in the bottom panel. \ The variable $z$ represents the \emph {imbalance function}.}
\end{figure}
\end{center}

In the case of cubic nonlinearity a family on nonlinear stationary states bifurcates from the linear stationary state when the adimensional nonlinear parameter $\eta$, associated with the strength $\epsilon$ of the nonlinear perturbation by (\ref {scala}), assumes the value $\eta^\star$ given by equation (\ref {formula11}). \ This nonlinear ground state branch consists of states having the same symmetry of the linear state, and tipically we observe also an exchange of the stability properties. \ The linear stationary state is stable for $\eta$ less than the value $\eta^\star$, and for $\eta$ larger than $\eta^\star$ the linear stationary state becomes unstable and the new asymmetrical states are stable: that is we have the usual picture of a pitch-fork bifurcation as in Figure \ref {Fig1} top-panel where the variable $z$ belongeing to  the interval $[-1,+1]$ represents the \emph {imbalance function}. \ Imbalance function, defined in equation (\ref {formula9}), is related to the position of mean value of the stationary state: when $z=0$ the state in invariant (up to a phase term) with respect to the symmetry of the double well potential; in contrast, when $z$ takes the end-point values $z=\pm 1$ then the state is fully localized inside on just one well (conventionally the right-hand side one for $z=+1$, and the left-hand side for $z=-1$). 

However, we should remark that the picture of Figure \ref {Fig1} top-panel still holds true also for other values of the nonlinearity power, for instance for $\mu =2$ and $\mu=3$, but for higher values of this parameter we observe a rather different picture \cite {Sacchetti1}. \ In Figure \ref {Fig1} bottom-panel we consider the case of nonlinearity power $\mu =5$, in such a case a couple of new asymmetrical stationary states sharply appear as saddle points when $\eta$ is equal to a given value $\eta^+ \approx 4.41$; then, for increasing values of $\eta$, the two unstable solutions disappear at $\eta =\eta^\star =6.4$ showing an inverse pitch-fork bifurcation. \ Thus, for $\eta$ between the two values $\eta^+ \approx 4.41$ and $\eta^\star =6.4$ we observe the co-existence of three stable stationary states; one of them corresponds to the linear stationary state which has the same symmetry properties of the potential, while the others two are localized on only one well. 

In this paper we investigate the bifurcation picture of the stable stationary states of equation (\ref {formula1})  for any positive value of the nonlinearity power $\mu$. \ In particular, in the semiclassical limit (or, equivalently, in the limit of large distance between the wells) we'll see that the simple pitch-fork bifurcation as in Figure \ref {Fig1} top-panel always occurs when the power $\mu $ is less than a critical value $\mu_{threshold}$, and the couple of saddle points with an inverse pitch-fork bifurcation as if Figure \ref {Fig1} bottom-panel always appears when the power $\mu $ is larger than $\mu_{threshold}$. \ The remarkable fact is that such a critical value $\mu_{threshold}$ is an \emph {universal critical power}, in the sense that it does not depends on the shape of the double well potential and it does not depend on the spatial dimension. \ Such a critical vale is found to be equal to
\begin{eqnarray}
\mu_{threshold} = (3+\sqrt {13})/{2} \, . \label {formula2}
\end{eqnarray}

The linear Hamiltonian we consider 
\begin{eqnarray}
H_0 = - \frac {\hbar^2}{2m} \Delta + V , \label {formula3}
\end{eqnarray}
with a symmetric double well potential $V(x) = V[{\mathcal S} (x)]$, where ${\mathcal S} (x)$ is the symmetric spatial inversion with respect to a given hyperplane $\Pi$ of the Euclidean space $ {\mathbf R}^{n} $. \ The potential has two nondegenerate minima at $x=x_\pm $, $x_+ = {\mathcal S} (x_-)$, such that $V (x) >  V(x_\pm )$, $\forall x \in {\mathbf R}^n \setminus \{  x_\pm  \}$, and $\nabla V (x_\pm ) =0$ and $ \mbox { Hess } V (x_\pm  ) >0 $. \ If we consider the semiclassical limit of $\hbar$ small enough \cite {Sacchetti1,Bambusi}, or equivalently the limit of large distance between the two wells  \cite {Sacchetti2,Kirr}, then it is well known that the discrete spectrum of $H_0$ is given by a sequence of doublets. \ Let $\lambda_{\pm}$ be a doublet of non-degenerate eigenvalues ($\lambda_+ < \lambda_-$), for instance the lowest two eigenvalues of $H_0$, then there exists a positive constant $C>0$, independent of $\hbar$, such that 
\begin{eqnarray*}
\inf_{\lambda \in \sigma (H_0) - \{ \lambda_{\pm} \} } [\lambda - \lambda_{\pm} ] \ge C \hbar \, , 
\end{eqnarray*}
where $\sigma (H_0)$ is the spectrum of $H_0$. \ The \emph {splitting} between the two eigenvalues $\omega = \frac 12 (\lambda_{-}-\lambda_{+}) $ exponentially vanishes as $\hbar $ goes to zero \cite {Landau}. \ The normalized eigenvectors $\varphi_{\pm}$ associated to $\lambda_{\pm}$ are even and odd real-valued functions with respect to the hyperplane $\Pi$
\begin{eqnarray*}
\varphi_{\pm} [{\mathcal S} (x)] = \pm \varphi_{\pm} (x ). 
\end{eqnarray*}
The normalized right and left hand-side vectors 
\begin{eqnarray*}
\varphi_R ={(\varphi_+ + \varphi_-)}/{\sqrt {2}} \ \ \mbox { and } \ \ \varphi_L =  {(\varphi_+ - \varphi_-)}/{\sqrt {2}},
\end{eqnarray*}
usually named \emph {single-well states}, are localized on only one well and their supports practically don't overlap in the sense that 
\begin{eqnarray}
\max_{x\in {\mathbf R}^n} | \varphi_{R } (x) \varphi_{L} (x) | = {\mathcal O} \left ( e^{- C /\hbar } \right ) , \ \mbox { as } \hbar \to 0 \, , 
\label {formula4}
\end{eqnarray}
for some positive constant $C$. 

The time dynamics associated to the linear Hamiltonian (\ref {formula3}) is well studied \cite {Kroemer}: when the state $\psi$ is initially prepared on the space spanned by the two vectors $\varphi_{R,L}$, then it performs a beating motion between the two wells with beating period $T= {2\pi \hbar}/{\omega}$. \ Since the beating period $T$ plays the role of the unit of time then we rescale the time $\tau = {\omega t}/{\hbar }$; furthermore, we consider also the gauge choice $\psi (x,t ) \to  e^{i \Omega t /\hbar} \psi (x,t ) $, where $\Omega = ( \lambda_+ + \lambda_- )/2 $. \ Then equation (\ref {formula1}) takes the form 
\begin{eqnarray}
i \omega {\partial_\tau \psi} = \left [ H_0-\Omega \right ] \psi + \epsilon |\psi |^{2\mu} \psi \, 
\label {formula5}
\end{eqnarray}
where we apply the two-level approximation by restricting the wave-function $\psi$ to the space spanned by the two single well states $\varphi_{R,L}$: 
\begin{eqnarray}
\psi =  a_R  \varphi_R + a_L  \varphi_L  \, , \label {formula6}
\end{eqnarray}
where $a_R$ and $a_L$ are unknown complex-valued functions depending on the time $\tau$. \ Since
\begin{eqnarray*}
H_0 \psi = a_R \left [ \Omega \varphi_R - \omega \varphi_L \right ] + a_L \left [ -\omega \varphi_R + \Omega \varphi_L \right ] 
\end{eqnarray*}
then, by substituting (\ref {formula6}) in (\ref {formula5}) and projecting the resulting equation onto the one-dimensional spaces spanned by the single-well states $\varphi_R$ and $\varphi_L$, it takes the form (hereafter $'$ denotes the derivative with respect to $\tau$)
\begin{eqnarray}
\left \{
\begin {array}{ll}
i \omega a_R' = - \omega a_L + {\epsilon} \langle \varphi_R , |\psi |^{2\mu} \psi \rangle \\ 
i \omega a_L' = - \omega a_R + {\epsilon} \langle \varphi_L , |\psi |^{2\mu} \psi \rangle  \label {formula7}
\end {array}
\right.
\end{eqnarray}
where $\langle \cdot , \cdot \rangle$ denotes the scalar product in the Hilbert space $L^2 ({\mathbf R}^n)$. \ From (\ref {formula4}) and since $\varphi_{R} [{\mathcal S} ( x)] =  \varphi_{L} (x)$, then a straightforward calculation led us to the following result
\begin{eqnarray*}
\langle \varphi_{R} , |\psi |^{2\mu} \psi \rangle = c |a_R|^{2\mu} a_R + {\mathcal O} (e^{-C_R/\hbar } ) 
\end{eqnarray*}
and
\begin{eqnarray*}
\langle \varphi_{L} , |\psi |^{2\mu} \psi \rangle = c |a_L|^{2\mu} a_L + {\mathcal O} (e^{-C_L/\hbar } ) 
\end{eqnarray*}
for some positive constants $C_R$ and $C_L$, and where the constant $c$ is the same in both equations and it is given by
\begin{eqnarray*}
c = \langle \varphi_R , |\varphi_R |^{2\mu} \varphi_R \rangle  = \langle \varphi_L , |\varphi_L |^{2\mu} \varphi_L \rangle
\end{eqnarray*}
Thus, the two-level approximation (\ref {formula7}) takes the following form up to an exponentially small error as $\hbar$ goes to zero
\begin{eqnarray}
\left \{
\begin {array}{lcl} 
i a_R' &=& - a_L + \eta |a_R|^{2\mu} a_R \\ 
i a_L' &=& - a_R + \eta |a_L|^{2\mu} a_L 
\end {array}
\right.   \label {formula8}
\end{eqnarray}
where 
\begin{eqnarray}
\eta = c{\epsilon}/{\omega} \label {scala}
\end{eqnarray}
is an adimensional parameter which only depends on the strength $\epsilon$ of the nonlinear term and, by means of the constant $c$ and of the splitting $\omega$, on the shape of the double well potential.

We perform now the qualitative analysis of the two level approximation (\ref {formula8}) by looking for the stationary states and studying their dynamical stability/instability properties. \ To this end we assume, for the sake of definiteness, $\eta >0$ and let
\begin{eqnarray}
a_R = p e^{i \alpha} , \ a_L = q e^{i \beta}, \ z = p^2 - q^2 , \ \theta = \alpha - \beta \label {formula9}
\end{eqnarray}
where $p$ and $q$ are such that $p^2+q^2 =1$. \ The \emph {imbalance function} $z$ takes value within the interval $[-1,1]$ and its value is related to the interval of localization of the wave-function $\psi$. \ The phase $\theta$ is a torus variable with values in the interval $[0,2\pi )$. \ Then, (\ref {formula8}) takes the Hamiltonian  form
\begin{eqnarray}
\left \{
\begin {array}{lcl} 
\theta' &=& \partial_z  {\mathcal H} \\
z' &=& - \partial_\theta {\mathcal H}  
\end {array}
\right. \label {formula10}
\end{eqnarray}
with Hamiltonian 
\begin{eqnarray*}
{\mathcal H} = 2 \sqrt {1-z^2} \cos \theta - \eta \frac { ( {1+z} )^{\mu +1} + ( {1-z} )^{\mu +1} }{2^\mu (\mu +1)} \, , 
\end{eqnarray*}
which is a first integral of motion

Equation (\ref {formula10}) always has, respectively, symmetrical and antisymmetrical stationary solutions $(\theta_1 , z_1 )$ and $(\theta_2 , z_2)$, where $\theta_1=0$ and $\theta_2=\pi$ and where $z_{1}=z_2=0$. \ Furthermore, asymmetrical stationary solutions may respectively occur for $\theta=0$ and $\theta =\pi$ as solutions of equations $f_+ (z) =0$ and $f_- (z)=0$, where 
\begin{eqnarray*}
f_\pm (z) = \mp  {2z}/{\sqrt {1-z^2}} - \eta [ ( {1+z}  )^{\mu } - ( {1-z} )^{\mu } ]  2^{-\mu} \, .
\end{eqnarray*}
Since we have assumed $\eta >0$ then the derivative $\frac {df_+}{dz}$ takes only negative values for any $z \in [-1,+1]$ and thus $f_+ (z) =0 $ has only the solution $z=0$. \ On the other hand, equation $f_-(z)=0$ might have other solutions coming from a pitch-fork bifurcation of the stationary solution $z=0$ as we can see in Figure \ref {Fig1} top-panel for $\eta$ larger than the value 
\begin{eqnarray}
\eta^\star = \lim_{z\to 0} \eta (z) = 2^\mu /\mu \label {formula11}
\end{eqnarray}
where 
\begin{eqnarray*}
\eta (z) = {2^{\mu+1} z}/\left [ \sqrt {1-z^2} ( (1+z)^{\mu } -(1-z )^{\mu } )\right ]
\end{eqnarray*}
is obtained by solving equation $f_- (z)=0$ with respect to $\eta$.

In particular, in Figure \ref {Fig1} bottom-panel we observe that in the case of $\mu =5$ a couple of saddle points appears when $\eta \in (\eta^+ , \eta^\star )$, where 
\begin{eqnarray*}
g(z,\mu )= (z^2 \mu - z \mu + 1 ) (1+z )^\mu \, .
\end{eqnarray*}
For instance, we have that $\eta^+ = \sqrt {27/2} \approx 3.67$, for $\mu =4$, and $\eta^+ \approx 4.41$ for $\mu =5$.

We thus have a transition from the bifurcation picture as in the top panel of Figure \ref {Fig1} to the more complex bifurcation picture as in the bottom panel of Figure \ref {Fig1}, and the transition from the first picture to the second one occurs when the nonlinearity power $\mu$ is equal to a threshold value $\mu_{threshold}$ such that $\frac {d^2 \eta (z)}{dz^2} =0$ at $z=0$, that is the two saddle points will merge with the stationary solution $z=0$. \ Since 
\begin{eqnarray*}
\left. \frac {d^2 \eta (z)}{dz^2} \right |_{z=0} = \frac {2^\mu (3 \mu +1 - \mu^2)}{9\mu}
\end{eqnarray*}
then the threshold value is given by (\ref {formula2}). \ We may remark that this threshold in an \emph {universal} value since it does not depend on the parameters of the double well model and on the spatial dimension. 

\begin{center}
\begin{figure}
\includegraphics[height=5cm,width=7cm]{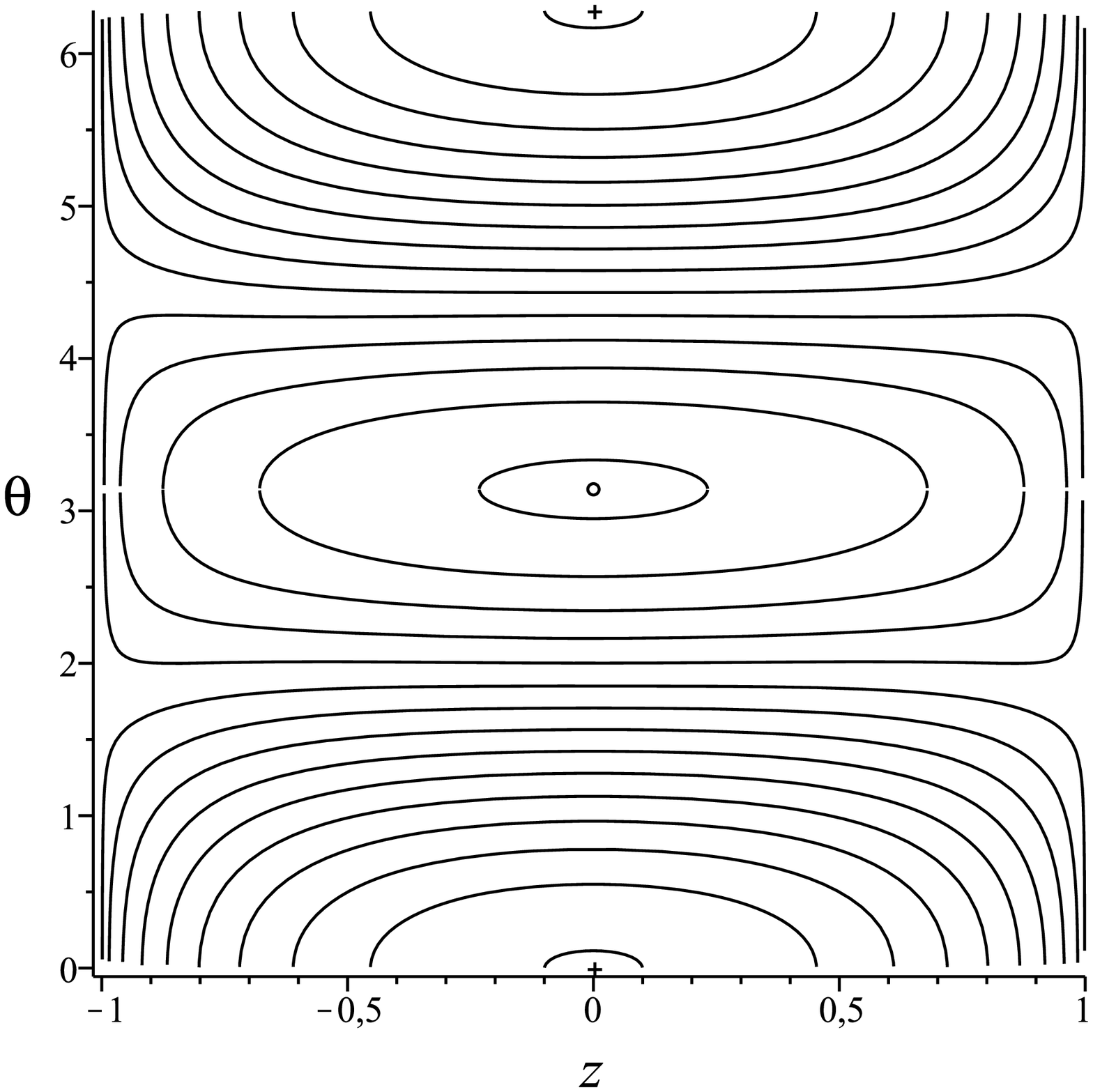}
\includegraphics[height=5cm,width=7cm]{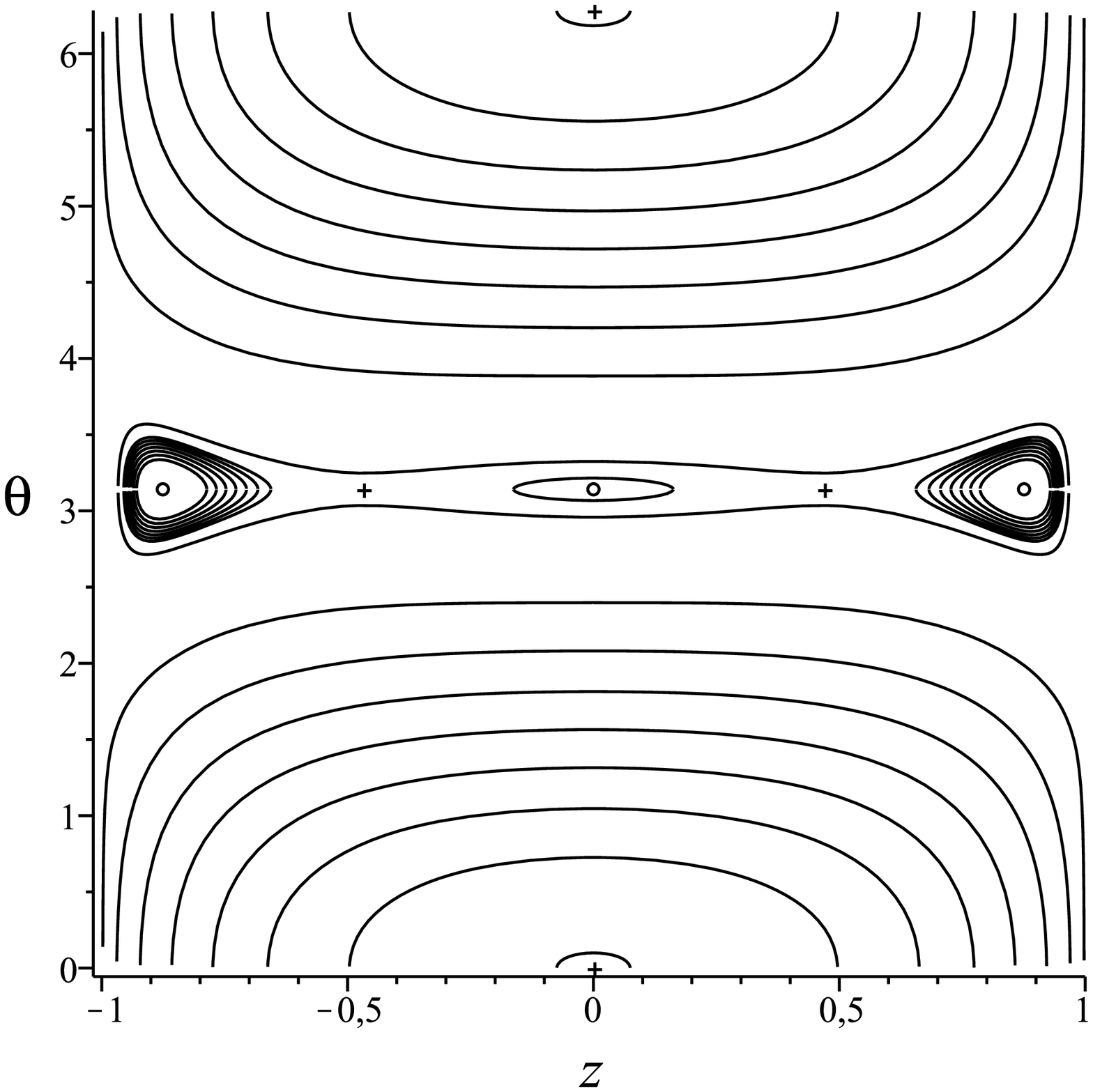}
\includegraphics[height=5cm,width=7cm]{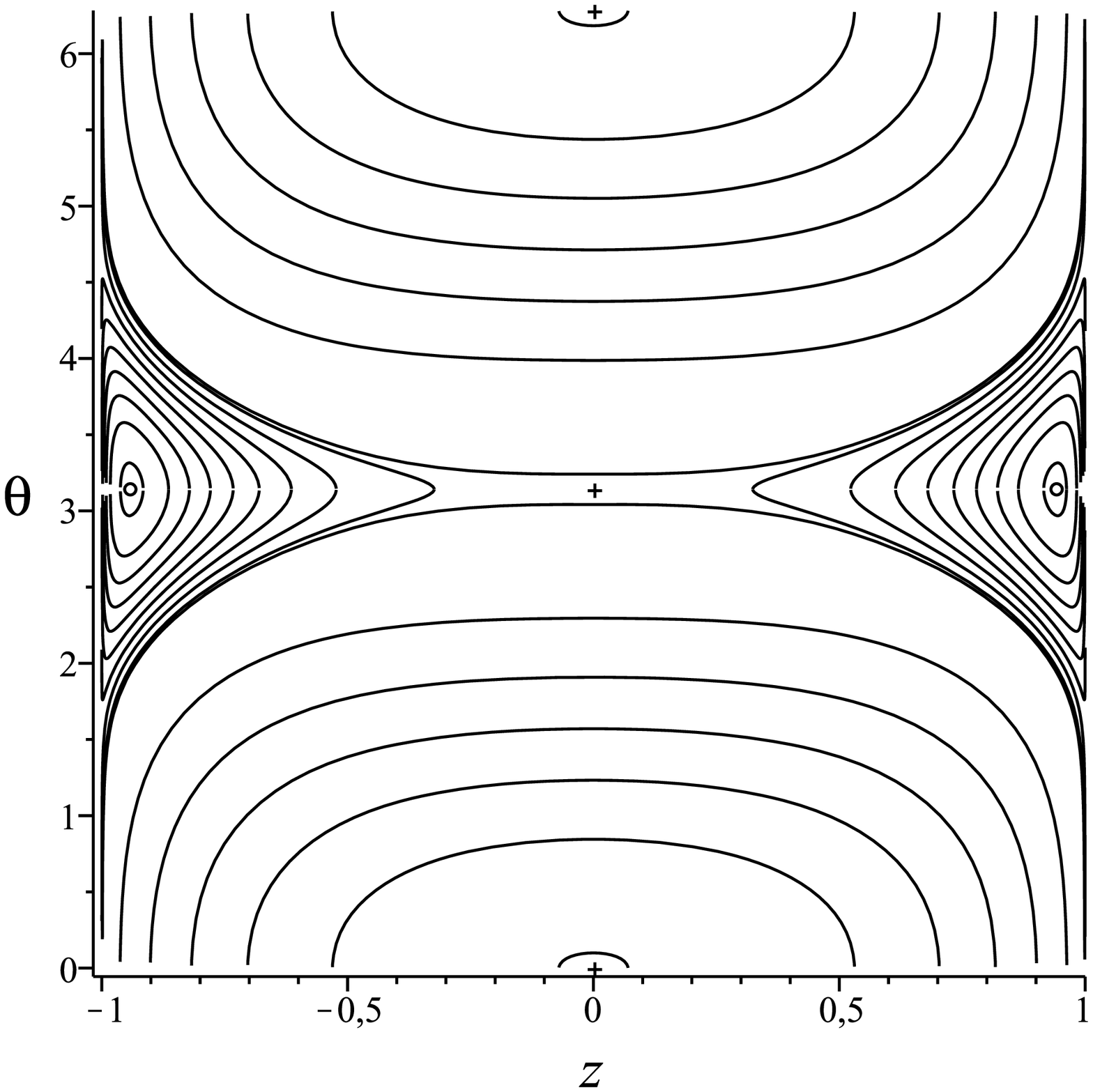}
\caption{\label {Fig2} Integral paths of the equation ${\mathcal H} (z,\theta) =E_n$ for some value of the energy $E_n$, where $\mu = 5$. \  Circle point corresponds to the stable stationary solutions, cross point corresponds to the unstable stationary solutions. \ Top panel correspond to $\eta =2$, less than $\eta^+ \approx 4.41$; middle panel correspond to $\eta =5$, which lies between the two values $\eta^+$ and $\eta^\star = 6.4$; bottom panel corresponds to $\eta =6.5$ larger than $\eta^\star=6.4$. \ For nonlinearity $\mu$ less than $\mu_{threshold}$ we only have the pictures of panels top and bottom, the picture of middle-panel does not occur.}
\end{figure}
\end{center}

The qualitative behavior of the solutions of equation (\ref {formula10}) is then studied by means of the conservation of the energy ${\mathcal H}$ as done, for instance, by \cite {Vardi} for cubic nonlinearity (where $\mu =1$). \ In Figure \ref {Fig2} we plot the integral paths of the equation ${\mathcal H} =E_n$ for some values of the energy $E_n$, where $\mu=5$. \ In the top panel where $\eta =2 < \eta^+ \approx 4.41$ we can only see closed curves corresponding to beating periodic motions between the two wells. \ In the middle panel, where $ \eta^+ \approx 4.41 < \eta =5 < \eta^\star=6.4$, we have three stable stationary solutions (circle points), two of them are localized on just one well and closed curves surrounding them correspond to periodic motions inside the well, without beating effect. \ In the bottom panel, where $ \eta^\star=6.4 < \eta =6.5$; we have two stable stationary solutions (circle points) localized on just one well and we don't observe a beating motion around the stationary solution at $z=0$.

In conclusion, in this paper we have proved the existence on an universal critical nonlinearity power (\ref {formula2}) for nonlinear Schr\"odinger equations with double well potential in the semiclassical limit. \ For nonlinearity power below this value we always observe a simple pitch-fork bifurcation phenomenon as the strength of the nonlinear perturbation increases, and the new asymmetrical stationary states gradually becomes localized on the single wells. \ In contrast, for nonlinearity power above (\ref {formula2}) we always observe a more complicate scenario: the appearance of a couple of saddle points where the asymmetrical unstable stationary solutions will merge with the stationary solution at $z=0$ drawing an inverse pitch-fork bifurcation as the strength of the nonlinear perturbation increases. \ The new physical relevant effect associated with such a new scenario is the sharply appearance of asymmetrical stationary solutions fully localized on the single wells.

\end {document}